# Ion hydration-controlled large osmotic power with arrays of Angstrom scale capillaries of vermiculite

Rathi Aparna[1], Dhal Biswabhusan[1], S S Sarath[1], Kalon Gopinadhan[1,2,3*]

[1]Department of Physics, Indian Institute of Technology Gandhinagar, Gujarat 382355, India

[2]Department of Materials Engineering, Indian Institute of Technology Gandhinagar, Gujarat 382355, India

[3]Lead contact

*Correspondence: gopinadhan.kalon@iitgn.ac.in

**SUMMARY**

**In the osmotic power generation field, reaching the industrial benchmark has been challenging because of the need for capillaries close to the sizes of ions and molecules. Here, we fabricated well-controlled 'along-the-capillary' membranes of Na-vermiculite with a capillary size of ∼ 5 Å. They exhibit 1600 times enhanced conductivity compared to commonly studied 'across-the-capillary' membranes. Interestingly, they show a very high cation selectivity of 0.83 for NaCl solutions, which resulted in large power densities of 9.6 W/m$^2$ and 12.2 W/m$^2$ at concentration gradients of 50 and 1000, respectively, at 296 K, for an unusually large membrane length of 100 μm. The power density shows an exponential increase with temperature, reaching 65.1 W/m$^2$ for a concentration gradient of 50 at 333 K. This markedly differs from the classical behavior and indicates the role of ion (de)hydration in enhancing power density, opening new possibilities for exploiting such membranes for energy harvesting applications.**

**2D material; Blue Energy; Vermiculite; Clay; Ion Hydration; Intercalation; Osmotic Power; Ion Transport; Cation-selective Membranes; Reverse Electrodialysis**

**INTRODUCTION**

The significant salt concentration difference at the seawater and river water interface is a clean source of enormous osmotic power of ∼ 2.4 TW[1]. This power is much larger than that solar and wind power produced together as of 2021. However, its extraction is difficult due to the absence of membranes with capillaries in the Angstrom scale. A minimum power density of 5 W/m$^2$ at a salt concentration gradient of 50 is required for practical applications. Recently, membrane-based reverse electrodialysis (RED) technology, which relies on the transport of salt ions, has emerged as a viable option thanks to the discovery of several thin two-dimensional (2D) materials. In the RED process, the power extracted by a membrane subjected to a salt concentration gradient is $\propto G.S^2$, where $G$ is the membrane conductance, and $S$ is the ion selectivity. This implies that osmotic power is maximum when the membrane exhibits high ionic conductance and good ion selectivity. 2D materials, due to their atomically small thickness, are a natural choice for high-conductance membranes[2]. Maximizing $S$ requires Angstrom scale fluidic channels with large surface charge density[3,4].

In this regard, J. Feng et al.[5] fabricated a single nanopore on a 1-layer $MoS_2$ membrane of thickness ∼ 0.67 nm. They reported an enormous power density of ∼ $10^6$ W/m$^2$ for a salt concentration gradient of 1000. Such power density mainly arises from the high conductance of the atomically thin membranes since the ion selectivity of K over Cl was poor, a mere 0.62 – 0.23. A large osmotic power density of ∼ $10^3$ W/m$^2$ was also reported in BN nanotubes of length 1 μm[6]. Though the power density looks very promising, they lack energy efficiency. Unfortunately, nanopores and nanotubes show broad size distributions making it even more challenging to extrapolate the diffusion properties from single to multiple pores/tubes. Recent studies, both simulation and experiment, find that the osmotic power does not linearly increase with the number of nanopores in a 2D sheet[7] or nanochannel length[8] due to increased ion concentration polarization in thin membranes. These at least hint that thin membranes are not ideal for high-efficiency osmotic power generators.

Several micron-thick membranes were extensively studied; however, in the pristine form, most of these membranes failed to achieve the required power density of 5 W/m$^2$, even for a salt concentration gradient of 500 between seawater and freshwater. Examples include MXenes[9], montmorillonite[10], graphene oxide[11], and metal-organic frameworks (MOFs)[12]. To improve the power density, researchers tried incorporating nanofibers (CNF) in GO [13], lithium intercalation in $MoS_2$ [14], treated GO with basic solutions[15], modified the vermiculite[16] or MXene[17] nanosheets to include pores, or vertically oriented the laminates[18,19]. This approach yielded promising results, yet at the realistic NaCl concentration gradient of 50, only a few membranes could reach power density above the benchmark. A few other studies reported surprisingly large power densities without any visible enhancement in conductance or selectivity. Our careful analysis concluded that the power densities were overestimated in those studies as the area considered was much smaller than the actual area of the membrane[17]. Few studies utilizing alumina nanochannels[20–22] and hydrogel membranes[23] reported power density values close to the bench mark. Few other studies demonstrated high power density at elevated temperatures, such as copper oxide[24], MXene/GO[25], and nacre-like silk-crosslinked[26] membranes. Among these, MXene/GO membranes displayed the most significant increase in power density of 112% at 343 K with respect to 298 K. A linear increase in diffusion coefficient, $D$, with temperature, $T$, as described by the Stokes-Einstein relation, $D = \mu k_B T/e$, readily explains this observation. Another critical requirement is the stability of these membranes in aqueous solutions to be considered for practical applications.

For our study, we picked vermiculite as the archetypal membrane material since it has a considerable surface charge density and small interlayer space. We recently successfully synthesized highly water-stable cation-intercalated vermiculite membranes[27]. In the present study, we specifically synthesized Na-intercalated vermiculite membranes to avoid any exchange of Na$^+$ from seawater. Additionally, we fabricated these membranes to transport salt ions along the interlayer spaces, which is defined as 'along the capillaries' transport. This is in contrast to routinely studied transport across the thickness, described as 'across the capillaries' transport[16,27,28]. With 'along-the-capillary' membrane, we observed a large power density of 9.6 W/m$^2$ and 12.2 W/m$^2$ at a NaCl concentration gradient of 50 and 1000, respectively, at 296 K for a membrane length of ∼ 100 μm, surpassing the industrial benchmark of 5 W/m$^2$. Interestingly, the power density displayed an exponential temperature dependence and increased by ∼ 578 % to 65.1 W/m$^2$ for a NaCl concentration gradient of 50 at 333 K. To our knowledge, this power density surpasses previous reports.

**RESULTS AND DISCUSSION**

**Synthesis of vermiculite and its characterization**

The sodium intercalated vermiculite (Na-V) membrane was fabricated with a procedure described in our previous work[27]. In short, the vermiculite crystals were intercalated with Li$^+$ ions and then sonicated in water. The vermiculite suspension in water has negatively charged nanosheets, as evidenced from the measured zeta potential of $\sim$ -45 mV. The membranes were prepared on PVDF support with the help of vacuum filtration and stabilized with NaCl. These micron-thick membranes naturally peeled off from the PVDF support, and we used these free-standing membranes for all of our studies (Inset Figure 1A). The X-ray diffraction (XRD) (GA-XRD, Rigaku SmartLab 9KW) data of the sample shows an intense peak at 2θ = 6.05° (Full width at half maximum = 0.77°), which translates into an interlayer spacing of 14.6 Å (Figure 1A and Figure S1). The net available space for the fluidic transport is $\sim$ 5.0 Å, after subtracting the space occupied by silicates of 9.6 Å. The XRD data also shows higher order diffraction peaks which can be described by Bragg's law, $n\lambda = 2d\sin\theta$, where the order of diffraction 'n' ranges from 1 to 5. The observation of higher order diffraction implies the highest quality of the laminate structure, presumably important for the demonstration of 'along-the-capillary' transport. We used an SEM (JEOL JSM-7900F) cross-section image to estimate the thickness of the membrane, which is $\sim$ 4 μm (Figure 1B).

**Fabrication of "along-the-capillary" transport samples**

We fabricated several devices with lengths, *L,* from 15 mm to 100 μm (Note S1). A strip of vermiculite membrane was encapsulated between two acrylic sheets with epoxy (Loctite Stycast 1266). This ensured that the only path for the ion transport is along the membrane's capillaries. The longer side was polished further using P1000 emery sandpaper to get the desired membrane length. The obtained sample was then glued onto an acrylic holder with a pre-fabricated hole of size 4 mm x 2 mm (Figure 1C). With effort, it was possible to reduce the length to tens of micrometers, though we limited our discussion to devices of length ≥ 100 μm. We used an optical microscope to precisely estimate the final length of the membrane (inset of Figure 1C).

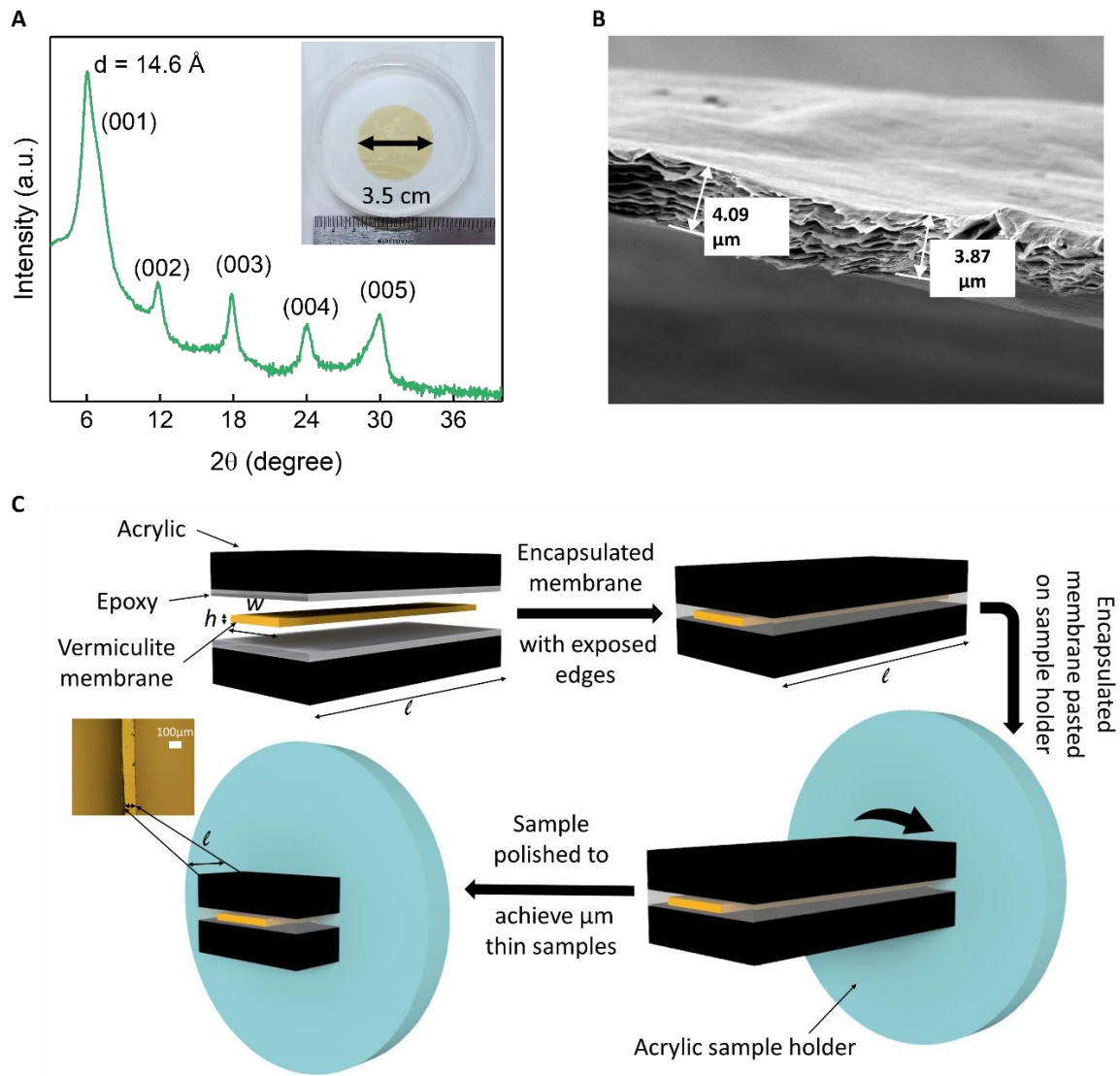

**Figure 1. Membrane characterization and sample preparation.**

(A) X-ray diffraction data of Na-V membranes, with an interlayer distance of 14.6 Å, estimated from the most intense first-order diffraction peak. Higher-order diffraction peaks are also clearly visible in the data. Inset: A free-standing membrane with a diameter of 3.5 cm in water.

(B) Cross-sectional SEM image of Na-V membrane, which provides a thickness estimate of ∼ 4 µm.

(C) Sample fabrication steps for 'along-the-capillary' transport devices. Inset: Optical microscope image of the acrylic encapsulating the membrane, which measures a membrane length of ∼ 100 µm. In the 'along-the-capillary' devices, this becomes the transport length.

**Ionic transport to understand the behavior of the membrane**

The membrane, along with the acrylic support, is placed in between two reservoirs made of PEEK (polyether ether ketone). We measured the ionic conductivity of the membrane using aqueous NaCl solutions with the same molar concentration, *C,* in both reservoirs. All the experiments were conducted at the room temperature of 296 K. We used a pair of home-made Ag/AgCl electrodes to convert ionic current to electronic and vice versa, which are connected to a source-measure unit (Keithley-Tektronix 2614B) and LabVIEW software. The schematic of the measurement is shown in the inset of Figure 2A. We applied a voltage, *V,* and measured the ionic current, *I,* that is transported through the membrane at several NaCl concentrations from $10^{-6}$ M to 1 M. The measured *I-V* characteristics were linear (Figure 2A) up to an applied voltage of 200 mV. The ionic conductance, *G*, is estimated from the slope of the *I-V* curves. We first measured *G* through a reference sample, which is an acrylic sheet with an open hole, and observed that *G* is proportional to *C*, as expected for the bulk transport[29]. We then measured the conductance of 'along-the-capillary' membrane at several concentrations of NaCl. For concentrations < $10^{-1}$ M, the magnitude of conductance is higher than that expected from the bulk transport and shows a sub-linear dependence on concentration. Our measured data is markedly different from the constant conductance expected from the surface charge at low concentrations. A power law fitting, $G \propto C^{\alpha}$, provides the value of the exponent, α, as 0.7. This probably suggests the role of hydration size and ion-ion interactions in determining the transport. Additionally, our measurements show a near-saturated ionic conductance at concentrations $>10^{-1}$ M (Figure 2B), quite similar to our previous study of ion transport 'across-the-capillary' in vermiculite membranes[27]. The conductance saturation probably indicates the restricted movement of ions due to the comparable size of our channels with the hydration size. The conductivity of ions for membrane 'along-the-capillary' is 1600-fold larger than 'across-the-capillary' membrane (Inset Figure 2B, Note S2, Table S1). This is plausibly arising from the well-defined capillaries along the interlayer direction than across the interlayer. It is worth mentioning here that van der Waals layered structure intrinsically possesses large anisotropic properties.

**Diffusion study to determine membrane's ion selectivity**

To understand the capability of these membranes for osmotic power generation, we performed drift-diffusion experiments at a temperature of 296 K. The experimental setup involved exposing the membrane to different concentration gradients, Δ of NaCl, as depicted in an inset of Figure 2C. Specifically, the concentration of NaCl on the feed reservoir ($C_H$) was kept constant at 1 M, while the concentration on the permeate reservoir ($C_L$) was systematically varied from 1 mM to 1 M. In the absence of any applied voltage, the concentration gradient drives a current through the membrane. We observed a negative diffusion current as shown in Fig. 2C, which is an indication of the preferential transport of cations. The voltage required to neutralize this current is denoted as diffusion potential, $V_{diff}$. The diffusion potential is recorded for several concentration gradients, Δ as shown in Figure 2D. This allowed us to estimate the selectivity, S, of the membrane using the Nernst equation (1) as,

$$V_{diff} = S \frac{k_B T}{ze} \ln\left(\frac{C_H}{C_L}\right) \quad \cdots\cdots\cdots (1)$$

Where, S = $t_+$ - $t_-$. Here, $t_+$ and $t_-$ denote the transport numbers for cations and anions, respectively, and $t_+$ + $t_-$ = 1, $k_B$ is the Boltzmann constant, $T$ is the temperature (296 K), $z$ = 1, $e$ is the elementary charge. For an ideal cation-selective membrane, $t_+$ = 1. We analyzed the variation of $V_{diff}$ with Δ (= $C_H/C_L$) using Eqn.1, providing a cation transport number, $t_+$ of 0.92 (S = 0.83). This suggests that Na-stabilized vermiculite membranes are cation-selective, a property required to generate large osmotic energy. We note there is a slight decrease in diffusion potential at Δ = $10^3$. This can be related to the increased boundary layer resistance[7], when the permeate concentration ($C_L$) is < 10 mM with a fixed feed concentration ($C_H$). Notably, the cation selectivity of 0.83 is much larger than 'across-the-capillary' membranes, which is only 0.64 (Figure S2). Knowing the cation transport number, $t_+$, we estimated the electrochemical energy conversion efficiency of our 'along-the-capillary' membranes[28,30] using the expression,

$$\eta_{max} = \frac{(2t_+ - 1)^2}{2} \times 100\% \quad \dots\dots\dots (2)$$

The energy efficiency is approximately 34%, which is quite large given that the theoretical maximum conversion efficiency of a perfectly selective membrane is only 50%.

We wanted to check if the most commonly used KCl can give a better power density than NaCl. For this measurement, we used K-intercalated vermiculite (K-V) membranes. With KCl, we observed a very high selectivity, S of 0.95 ($t_+$ = 0.97), compared to NaCl, where the latter has S = 0.83 ($t_+$ = 0.92) (Figure S3A). However, KCl conductance through the K-V membrane was relatively low, most likely arising from the smaller interlayer spacing of K-V[27] compared to Na-V. This resulted in a power density of 10 - 15 % less compared to the Na-V membrane with NaCl (Figure S3B); however, the energy conversion efficiency of K-V with KCl was close to 45 %.

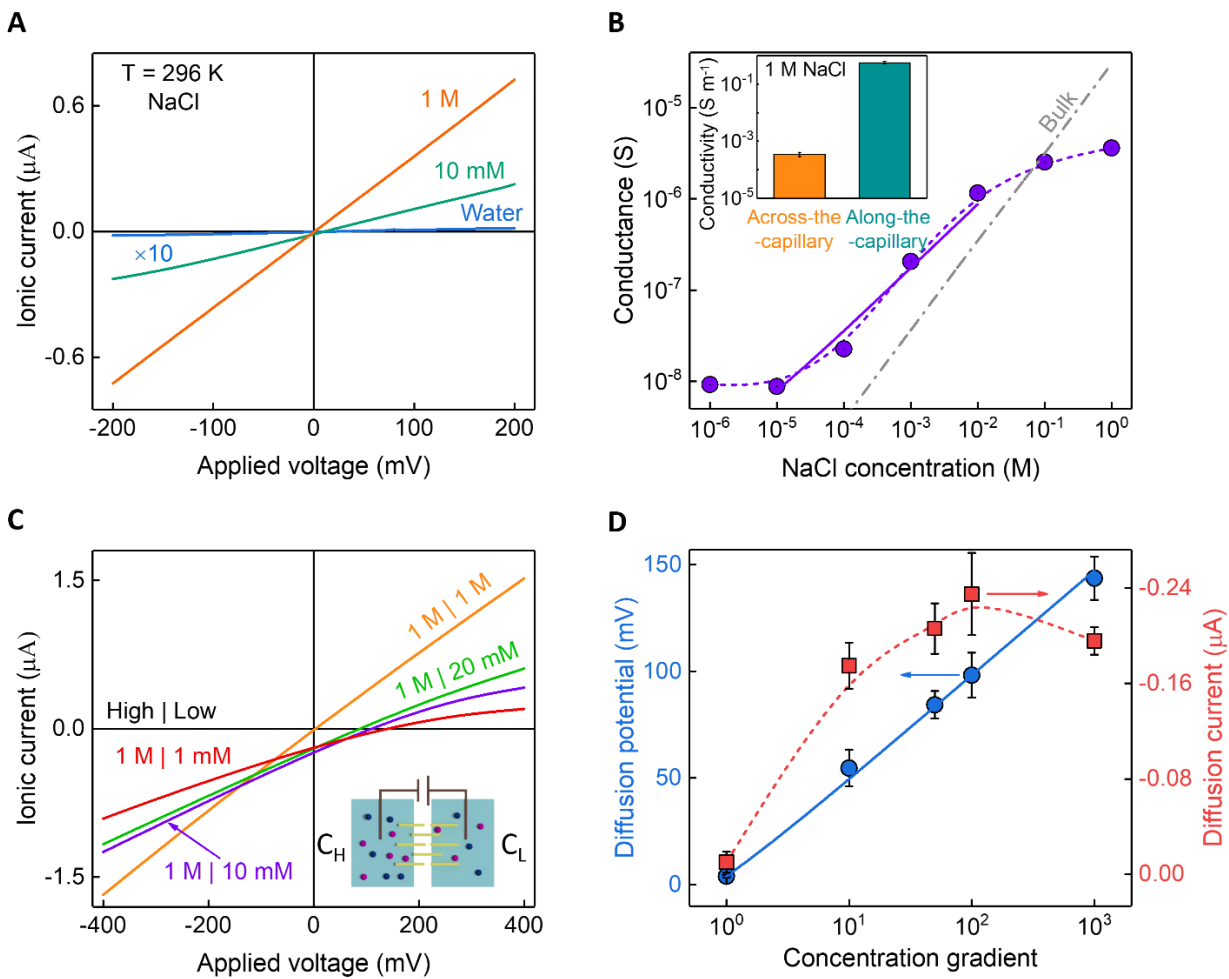

**Figure 2. Ion transport through Na-V membrane.**

(A) Representative *I-V* characteristics of ~ 4 μm thick 'along-the-capillary' membrane; $L \sim 100$ μm, $w \sim 132$ μm, for few concentrations of NaCl solutions. For clarity, the curve for water with a molar concentration of $10^{-6}$ M is magnified by 10 times. Inset: Schematic of the conductance measurement setup.

(B) The variation of ionic conductance with molar concentration of NaCl. The dotted violet curve is a guide for the eye, while the solid line is a fit. The bulk conductance of NaCl (membrane area = 528 μm$^2$, length = 100 μm) is represented by the grey dotted line[29]. Inset: Comparison of 1M NaCl conductivity of 'across-the-capillary' and 'along-the-capillary' devices. The error bars indicate the standard deviation of data for 3 samples.

(C) *I-V* characteristics for several concentration gradients of NaCl, indicating the diffusion potential and diffusion current. The redox potential of the Ag/AgCl electrodes is already subtracted from the data. Inset: Schematic of 'along-the-capillary' diffusion measurement setup.

(D) The variation of diffusion potential and current as a function of NaCl concentration gradient. The diffusion potential is fitted using the Nernst equation shown by the blue solid line. The red dotted line for diffusion current is a guide for the eye. Error bars indicate standard deviation of three independent measurements for diffusion potential and current.

**Variation of power density with length of the membrane**

It is interesting to know how the length, *L*, of the membrane affects the power density. We started with an *L* of 15 mm, the membrane exhibited a selectivity of 0.83 and a power density of 0.05 W/m² for a NaCl concentration gradient of 50. However, when *L* was reduced from millimeters to micrometers, a significant improvement in power density was observed. For example, the power density increased 175 times from ∼ 0.05 W/m² to 9.6 W/m² at Δ = 50 (Figure 3A) when the length decreased by 150 times, i.e., from 15 mm to 100 μm. Interestingly, the cation selectivity was not much affected and remained at 0.83 (Figure S4). This points to the role of membrane internal resistance, $R = \frac{\rho L}{A}$, and the channel length, *L*, that ions have to travel across the membrane. We observed that *R* is linearly proportional to *L* of the membrane along the transport direction (Figure S5). The ions 'along-the-capillary' shows better cation selectivity as the transport is influenced by negative surface charges of the walls and interestingly, the large number of parallel capillaries provide a high conductivity path. In comparison, the ion transport 'across-the-capillary' configuration is tortuous, with a large resistance to the movement of ions and hence a smaller conductivity (Inset of Figure 2B and Figure S6).

To estimate how much power this device delivers to an external load, we evaluated the power from the measured *I-V* curves as well as by testing the device under several external load resistances. The maximum output power density was calculated using the relation, power density, P = $V_{diff}^2$/(4R × Area), where the Area = 528 μm². A power density of 9.6 W/m² was recorded at an external load resistance, $R_L$ of 291 kΩ, for a NaCl concentration gradient, Δ of 50 (Figure 3B-C and Figure S7). The load resistance matches with the device's internal resistance extracted from the *I-V* curves (Figure 2C). Similarly, for a Δ = 100, a power density of 11.7 W/m² was recorded at $R_L$ of 345 kΩ; for Δ = 1000, a maximum power density of 12.2 W/m² was extracted (Figure 3B-C).

**The variation of osmotic power with solution pH**

Next, we wanted to check whether the pH of the electrolyte (Note S3) affects the osmotic power. We observe that the power density shows an increase from 6.4 to 13.1 W/m² at Δ = 50 when pH increases from 3 to 11, which is approximately a factor of 2 (Figure 3D). Our estimated cation selectivity also shows an increase from 0.61 to 0.93 when the solution pH increases from 3 to 11 (Figure 3D). However, *G*, shows a slight reduction when pH increases from 3 to 11 (Figure S8A). From the relation, power ∝ *G.S²*, it is clear that the increase in power density is primarily due to the change in ion selectivity. Vermiculite has OH⁻ groups present on tetrahedral and octahedral sheets; with a change in the environmental pH, these groups might experience protonation/deprotonation, causing some temporary changes in the surface charge of the membrane, which is reflected in the selectivity of the membrane for Na⁺ (Ref.[31]). At larger pH, the vermiculite layer charge increases as evident from the selectivity, which presumably reduces the interlayer spacing. This would imply a larger hydration barrier for the movement of ions at higher pH values. Our temperature-dependent conductance measurements suggest that the activation energy indeed increased from 33.3 kJ mol$^{-1}$ to 36.9 kJ mol$^{-1}$ with pH (Figure S8B). The pH study also provides a measure of the stability of these membranes in both mild acidic and basic conditions.

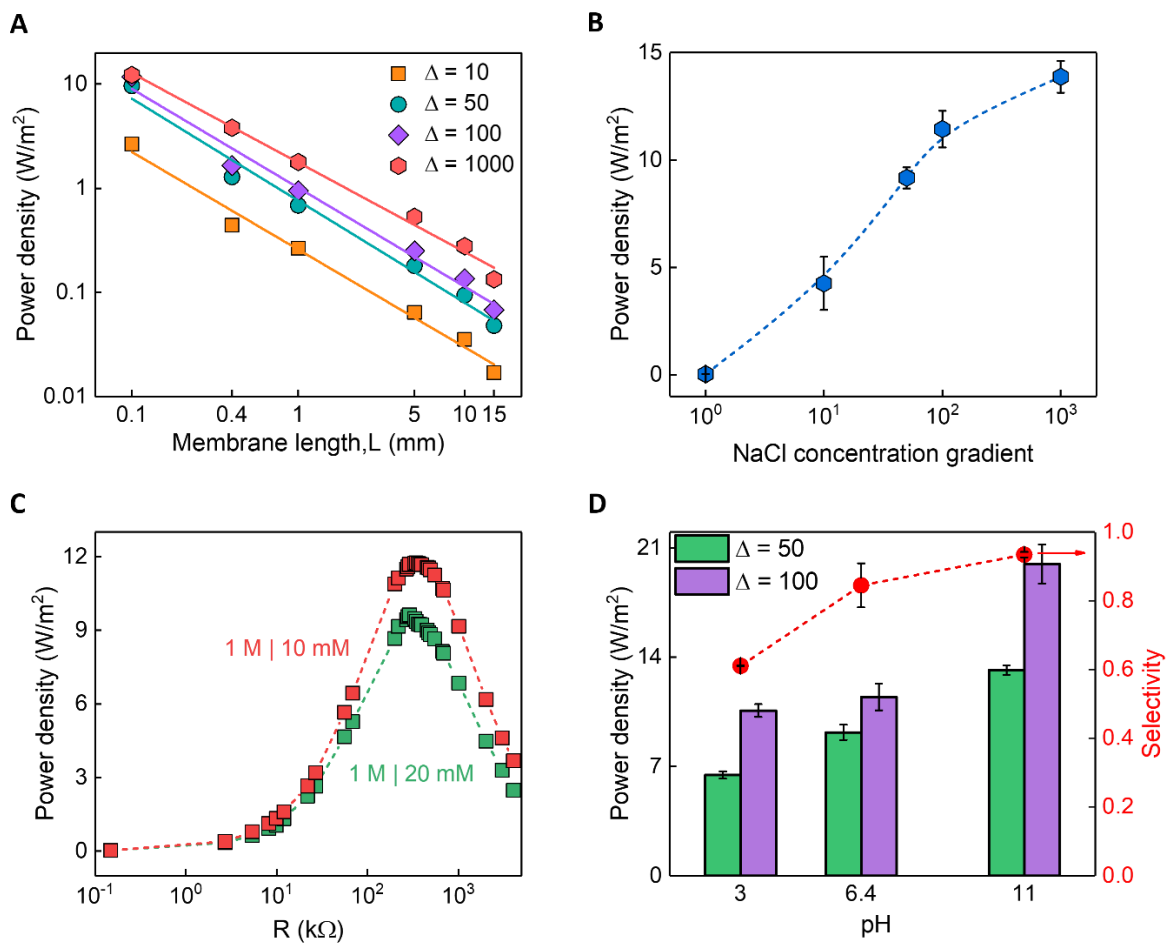

**Figure 3. Factors contributing to the change in power density of Na-V membrane.**

(A) Power density as a function of membrane length, where shorter capillaries display larger power densities crossing the industrial benchmark of 5 W/m$^2$ at room temperature. The solid lines are a linear fit to the data.

(B) Variation of power density with concentration gradient where 1 M NaCl was fixed on one side of the membrane and on the other side, the concentration was varied from 1 mM to 1 M at T = 296 K.

(C) Power density variation as a function of load resistance for NaCl concentration gradients of 50 and 100 at room temperature of 296 K. Where Δ = 10$^2$ provides a maximum power density of 11.7 W/m$^2$ with an external load resistance (R$_L$) of 345 kΩ, and at Δ = 50, we get a power density of 9.6 W/m$^2$ with R$_L$ = 291 kΩ.

(D) Variation of power density (Left Y-axis) and cation selectivity (Right Y-axis) with pH at Δ = 50 and 100 for T = 296 K. In fig. B,C and D, the membrane length is ≈ 100 μm. All the dotted lines are guides for the eye. Error bars indicate the standard deviation of three independent measurements.

**Variation of power density with temperature**

The remarkable thermal stability of vermiculite[32] made us investigate the temperature dependence of power density. We placed the vermiculite membrane-contained PEEK cell on a hot plate and gradually varied the temperature of the saline solutions in both reservoirs (Figure S9A). It should be noted that there is no temperature gradient across the membrane, and all the measurements were done when thermal equilibrium was achieved. We performed drift and diffusion measurements at several temperatures and estimated conductance and diffusion potential at each temperature (Note S4, Figure S9B). The estimated power density shows an exponential increase with temperature (Inset Figure 4A). Over this temperature range, the cation selectivity only slightly increased by ∼ 10%. For a NaCl concentration gradient of 50, the power density was 9.6 W/m² at 296 K, which increased to 65.1 W/m² at 333 K (Figure 4A, Table S2). The Stokes-Einstein relation between diffusion coefficient and temperature cannot simply explain this increase. In the bulk case, both diffusion coefficient and mobility are linearly dependent on temperature (Note S5, Table S3); however, in our membranes, an exponential increase in the mobility of cations with temperature is observed (Figure 4B, Figure S10). At higher temperatures, the cations are more prone to lose their hydration shell, especially with smaller cations such as Na$^+$. Hence, a reduction in the size of the hydrated cation leads to an increase in its mobility and an increased number of water molecules in the solution[33]. Apart from this, in the case of hydrated clay, with a rise in temperature, the vibrations of interlayer cations also increase, causing an expansion of the interlayer spaces leading to an increase in the diffusion coefficient of both water molecules and cations[34]. Moreover, our membranes displayed a sub-linear ionic conductance with concentration which suggests the importance of hydration sizes in ionic transport. All these suggest that the increase in ionic conductance at higher temperatures can be accounted for by the ion de(hydration) effects. The energy barrier for ion transport through Na-V is calculated using the Arrhenius equation:

$$\sigma = \sigma_o \, exp\left(-\frac{E_A}{RT}\right) \quad \ldots\ldots\ldots\ldots (3)$$

Where $\sigma$ is the ionic conductivity, $\sigma_o$ is the Arrhenius constant, $E_A$ is the activation energy, $R$ is the universal gas constant, and $T$ is the temperature. The Arrhenius plot of conductivity for 'along-the-capillary' and 'across-the-capillary' membranes provide activation energy values of 36.0 kJ mol$^{-1}$ and 37.5 kJ mol$^{-1}$, respectively (Figure 4C). A similar activation energy suggests that in both transport configurations, ions encounter hydration barrier in capillaries, however, in the case of 'across-the-capillary' configuration, the presence of defects and a longer path probably slightly increases the energy barrier.

To understand this anomalous temperature dependence of power density, we performed COMSOL Multiphysics simulations as described in Note S6 (Table S4). We modelled the ionic mobility as:

$$\mu_\pm \propto exp\left(-\frac{E_A}{RT}\right) \quad \ldots\ldots\ldots\ldots (4)$$

With this, we could match the experimental diffusion I-V curves, which further confirms the ion (de)hydration behavior with increase in temperature (Figure S11, 12).

**Membrane's performance and stability**

To test the stability of these membranes over long-term operation, we measured the diffusion characteristics at regular intervals. Under ambient conditions, the membrane exhibits stable diffusion potential and power density when operated for more than 60 hours (Figure 4D). The fluctuations in the data can be accounted for by the evaporation of the solution over days and slight variations in laboratory temperature. However, the device maintained a power density of > 8.5 W/m$^2$ at a NaCl concentration gradient of 50 for more than 60 hours, implying these membranes' robustness under ambient conditions. To demonstrate its usefulness, we increased the membrane area to ∼ 0.4 mm$^2$, stacked multiple membranes using epoxy (Figure S13A), connected six such cells and operated under a NaCl concentration gradient of 150 (3 M | 0.02 M). We obtained an output voltage of 1.66 V (Figure S13B) and a current of 10 µA. This setup was able to power up red and yellow light-emitting diodes (LEDs) ($V_{working}$ = 1.6 V) (Figure S13C, Video S1). We calculated the power density for these stacked membranes to infer any losses due to increased area. These 0.4 mm$^2$ area membranes gave a power density of ∼ 3.0 W/m$^2$ for a concentration gradient 150. The membrane length used in this experiment is ∼ 500 µm, larger than the shortest devices, to avoid defects at the multiple epoxy-membrane interfaces during polishing. The estimated power density is comparable to non-stacked membranes (Figure 3A), indicating no loss of power. With better control of the epoxy-membrane interface by fine-tuning the epoxy process parameters, short-channel devices (L < 500 µm) are also feasible to scale up similarly to long-channel devices. Although our non-stacked membranes are of small area compared to literature reports provided in Table S5, the overall area can be further improved with thicker and wider membranes and stacking.

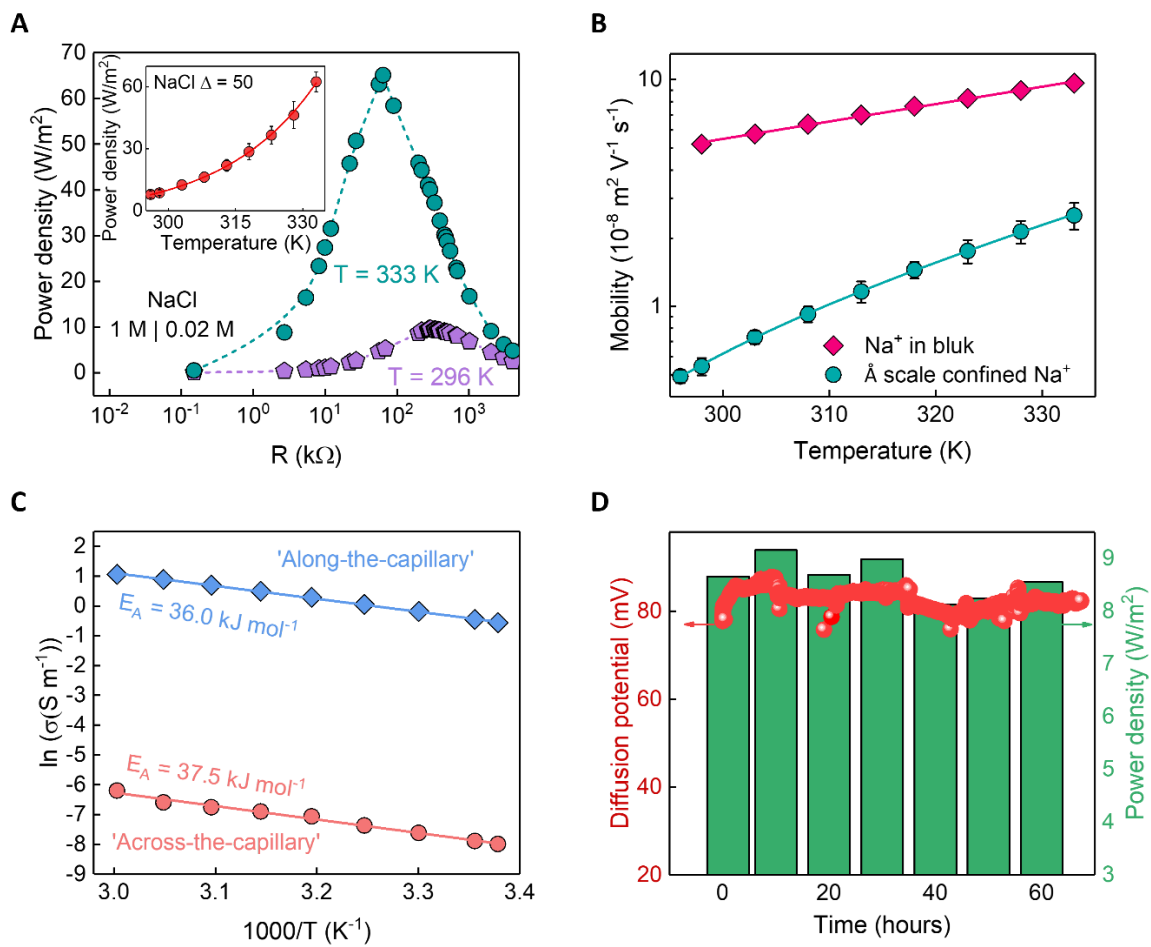

**Figure 4. Temperature-dependent studies.**

(A) Power density as a function of external load resistance for T = 296 K and T = 333 K (L ≈ 100 µm), showing ∼ 578 % times increase in power density while the membrane's resistance decreases from 291 kΩ to 65 kΩ with temperature. The dotted lines are guides for the eye. Inset: Power density grows exponentially as a function of temperature, reaching a value as high as 65.1 W/m² at T= 333 K from 9.6 W/m² at 296 K for a NaCl concentration gradient, Δ = 50. The solid line in the inset is the exponential fit to the data.

(B) The mobility of Na⁺ across-the-capillary is plotted as a function of temperature. The green solid line is the exponential fit. Bulk mobility values of hydrated Na⁺, taken from literature[35], at different temperatures, is also plotted. The pink solid line is the linear fit to the data.

(C) The Arrhenius plot of ionic conductivity of 1 M NaCl for both 'along-the-capillary' and 'across-the-capillary' membranes. The solid lines represent the linear fit to the data.

(D) Stability test of the membrane diffusion potential (Left Y-axis) and power density (Right Y-axis) as a function of time for NaCl Δ = 50 (L ≈ 100 µm). The error bars represent the standard deviation of three independent measurements.

Our device fabrication and experiments were mainly focused on improving the conductance, $G$ and ion selectivity, $S$. From the relation, $G = \sigma A/L$, it is clear that for unit membrane area A, higher $G$ is possible only if the salt conductivity, $\sigma$, is higher. The parallel capillaries provided higher $\sigma$ at sea salt concentrations in contrast to 'across-the-capillary' transport, which explains our results. Our attempt to pick vermiculite capillaries served two purposes. The inherent structure of vermiculite supports large surface charge densities, which lead to higher ion selectivity. The comparable size of our capillaries and the Debye screening length ensured high ion selectivity even at 1 M concentration. It is to note that the Debye length at 1 M concentration is 0.3 nm.

The hydration effects are found to be crucial to enhance the power density many folds. If the channel sizes are much smaller than the hydration size, this will result in smaller ionic currents, and hence the achievable power will be lower, which is not desirable. If the channels are much larger than the hydration sizes, the hydration effects will weaken, resulting in a bulk-like temperature dependency on power. Therefore, it is desirable to have membranes with channel size close to hydration size to observe a similar power density variation with temperature. This effect can be explored in MXenes, and Al-stabilized GO, which could be potential candidates for future research. In the clay's family, montmorillonite, bentonite, talc, kaolinite, and gypsum could be explored for these applications. Even though there have been many studies to enhance the osmotic power density but most of them fail to meet the industry bench mark with NaCl solutions. The decreased mobility of $Na^+$ as compared to $K^+$ is one of the reasons that limits these studies to solutions like KCl[5,6]. The water and thermal stability of vermiculite along with its low cost and environmentally friendly makes it stand out from other materials. A comparison of the power density generated by our membranes and several other studies at a NaCl concentration gradient of 50 has been provided in Table S5.

We demonstrated that 100 µm long membranes could generate high power densities. Apart from the mechanical polishing method discussed in this work, alternative techniques like ion beam thinning (Ion milling/ FIB), ultramicrotomy, and electropolishing can also be employed to achieve thinner samples. A better power density is expected if length is reduced below 100 µm. However, reducing the length to less than microns will likely not enhance the power density due to concentration polarization[8]. Solar concentrators can be used in our devices to provide higher temperatures, which is definitely a cheaper way to achieve higher power densities. Alternatively, these devices can also be operated near the outlet of air conditioners, where temperatures of 50-60 ℃ are readily available. We found these devices stable under several thermal cycles, with no conductance hysteresis under heating and cooling. The maximum temperature we used in this study is safe for the different components of the membrane device, such as epoxy. However, excessive heating beyond a specific limit can jeopardize the integrity of the membrane.

In conclusion, by transporting the ions along the interlayer space direction, we achieved high ion selectivity and conductance, which lead to high power density and energy efficiency. The angstrom-sized capillaries provided a hydration barrier for the ions, which the ions needed to overcome. Because of this, the ion mobility shows an exponential variation with temperature, resulting in an exponentially large power density. We reported an enormous power density of 65.1 W/m$^2$ at an accessible temperature of 333 K. We combined six such devices to demonstrate its ability to power up red and yellow LEDs. The shortest membrane that we tried was 100 µm long, however, this could be further scaled down with several techniques, promising even better power density. By choosing capillaries that are close to the hydration size of the ions, it is possible to control the ion transport with a (de)hydration barrier. The manipulation of ion (de)hydration characteristics provides another possibility to separate different ions from a mixture of ions. The success in this direction would primarily depend on our ability to fabricate angstrom scale capillary devices.

**EXPERIMENTAL PROCEDURES**

**Resource Availability**

**Lead Contact**

Further information and requests for resources should be directed to and will be fulfilled by the Lead Contact, Kalon Gopinadhan (gopinadhan.kalon@iitgn.ac.in).

**Materials Availability**

This study did not generate new unique materials.

**Synthesis of vermiculite membranes**

A two-step ion exchange was used where 100 mg of vermiculite crystals (Sigma Aldrich) were dispersed in 200 ml of saturated sodium chloride (NaCl) solution, heated for 24 hours at 100°C, and washed with water. The product is then dispersed in a 2 M lithium chloride (LiCl) solution and again heated for 24 hours, expanding interlayer spaces of vermiculite crystals. These crystals were then rigorously washed and dried to remove any residual salts. The resultant Li-V crystals were added to water with a 1 mg/ml concentration and sonicated for 40 minutes. The obtained suspension was centrifuged at 3000 rpm for 10 minutes to remove unexfoliated and multilayer flakes. The supernatant was used to make membranes via vacuum filtration on 0.22 µm pore size PVDF support. The fabricated Li-V membrane was unstable in water and hence stabilized by dipping in 1 M NaCl for 24 hours. The membrane was thoroughly washed to remove any excess salt that may be present on the surface and then used for "along-the-capillary" device preparation.

**Data and Code Availability**

Data generated from this study will be made available upon reasonable request to the lead author.

**SUPPLEMENTAL INFORMATION**

Document S1. Note S1-S6, Figure S1-S13 and Table S1-S5

**Video S1.** Demonstration of LED illumination with the help of osmotic power generated via our setup containing "along-the-capillary" Na-V membranes.

**ACKNOWLEDGMENTS**

This work was mainly funded by Science and Engineering Research Board (SERB), Government of India, through grant CRG/2019/002702 and also supported by MHRD STARS with grant no. MoE-STARS/STARS-1/405. R.A. acknowledges the Prime Minister Research Fellowship (PMRF) from the Ministry of Education, Government of India. We also acknowledge the financial support from DST-INAE with grant no. 2023/IN-TW/09. K.G. acknowledges the support of Kanchan and Harilal Doshi chair fund. The authors acknowledge the contribution of SEM facility from DST, Government of India (SR/FST/ET-I/2017/18) and IITGN central instrumentation facility.

**AUTHOR CONTRIBUTIONS**

**R.A.:** Investigation, methodology, visualization, validation, writing- original draft, writing – review & editing. **D.B.:** Software. **S.S.S.:** Validation. **K.G.:** Supervision, conceptualization, validation, methodology, writing – review & editing, funding acquisition, project administration.

**DECLARATION OF INTERESTS**

The authors declare no competing interests.

**SUPPLEMENTAL NOTES**

**Note S1. Sample preparation procedures.**

To achieve membrane lengths < 5 mm, the samples were encapsulated between two acrylic pieces with the help of epoxy and polished with emery paper P1000 to expose the inlet and outlet of membrane channels along the length. The obtained sample was then pasted on an acrylic holder with a 4 mm x 2 mm hole in the center and then polished until the desired length was achieved.

All the samples ≥ 5 mm in length were simply covered with epoxy, and with the help of a scalpel blade, the edges were cut to obtain the desired length. These samples were then attached to the reservoirs for further studies.

**Note S2. Enhancement in the conductivity of 'along-the-capillary' transport as compared to 'across-the-capillary' transport.**

To study the effect of 'along-the-capillary' transport and 'across-the-capillary' transport, we compared the conductivity of 1 M NaCl through the channels of Na-V membrane in both orientations. For 'along-the-capillary' transport, membrane with a cross-sectional area of 528 µm$^2$ and length of 100 µm was used, while for 'across-the-capillary' transport, the cross-sectional area was 1.6 mm$^2$, and the membrane thickness was 4 µm. The conductivity, as determined by the equation G = σA/*L*, exhibits a 1600-fold increase for 'along-the-capillary' transport conductivity due to less resistance experienced by the ions during the transport.

**Table S1**. Conductivity comparison between 'along-the-capillary' transport and 'across-the-capillary' transport.

| Temperature (K) | 'Along-the-capillary' conductivity (S m$^{-1}$) | 'Across-the-capillary' conductivity (S m$^{-1}$) | Conductivity ratio |
|---|---|---|---|
| 296 | 0.5687 | 3.3978 × 10$^{-4}$ | 1673 |
| 298 | 0.6348 | 3.7525 × 10$^{-4}$ | 1691 |
| 303 | 0.8323 | 4.9352 × 10$^{-4}$ | 1686 |
| 308 | 1.0520 | 6.3255 × 10$^{-4}$ | 1663 |
| 313 | 1.3362 | 8.5803 × 10$^{-4}$ | 1557 |
| 318 | 1.6379 | 0.0010 | 1637 |
| 323 | 2.0052 | 0.0011 | 1822 |
| 328 | 2.4195 | 0.0013 | 1861 |
| 333 | 2.8866 | 0.0020 | 1443 |

**Note S3. Preparation and calculations for pH solutions.**

To change the pH of the NaCl solution, 1 M HCl and 1 M NaOH were used as acid and base buffer, respectively. The following calculations were used to prepare 1 M NaCl and 0.02 M NaCl for pH 3 to acquire a concentration gradient of 50. A 30 µL solution of 1 M HCl was added to 5 mL of 1 M NaCl solution to obtain 1 M NaCl solution of pH 3. For the low concentration side, to prepare ∼ 0.2 M of a solution, we added 30 µL of 1 M HCl to 5 mL of 0.014 M NaCl. To finalize the volume and molarity of the mixing solutions, we used the equation: Molarity of the solution = n/V, where n is the total number of moles = (Molarity of NaCl × Volume of the NaCl) + (Molarity of buffer solution × Volume of the buffer solution), and V is the total volume of the solution = Volume of NaCl solution + Volume of buffer solution. The same procedure was used to prepare solutions with pH 11.

**Note S4. Variation of power density with respect to temperature.**

Since the power density, P = $V_{diff}^2/(4R \times Area)$, where $V_{diff} = V_0 - V_R$, is the membranes diffusion potential, $V_0$ is the zero-current potential, $V_R$ is the redox potential, R is the resistance (1/G), and Area = 528 µm². The diffusion potential and resistance values (1/G) obtained from *I-V* graphs at different temperatures are used to calculate the power density, which displays an exponential increase with temperature. The percentage change in power density can be calculated by $\frac{\Delta P}{P_{296\ K}} \times 100$ %, where $\Delta P = P_{333\ K} - P_{296\ K}$. We observed a 578 % increase in power density with a rise in temperature from 296 K to 333 K.

**Table S2.** The output power density of one of the L ≈ 100 µm membranes with a NaCl concentration gradient of 50 across a range of temperature from 296 K to 333 K, which shows a 578 % increase in power density for ΔT = 37 K.

| Temperature (K) | Measured potential, $V_o$ (V) | Redox potential, $V_R$ (V) | Diffusion potential, $V_{diff}$ (V) | Resistance (kΩ) | Power density (W/m²) |
|---|---|---|---|---|---|
| 296 | 0.177 | 0.099 | 0.077 | 291 | 9.6 |
| 298 | 0.178 | 0.100 | 0.078 | 270 | 10.3 |
| 303 | 0.182 | 0.102 | 0.080 | 227 | 13.3 |
| 308 | 0.186 | 0.103 | 0.082 | 180 | 17.6 |
| 313 | 0.191 | 0.105 | 0.086 | 141 | 24.7 |
| 318 | 0.194 | 0.107 | 0.087 | 115 | 30.9 |
| 323 | 0.199 | 0.109 | 0.090 | 94 | 40.6 |
| 328 | 0.204 | 0.110 | 0.094 | 78 | 53.4 |
| 333 | 0.207 | 0.112 | 0.095 | 65 | 65.1 |

**Note S5. Temperature dependence of mobility.**

The diffusion coefficient, D, of the bulk ions was calculated using the Nernst-Einstein equation:

$$D = \frac{RT}{|z|F^2}\Lambda_m^o \quad (1)$$

Where R is the universal gas constant, T is the temperature, $|z| = 1$, F is the Faraday constant, and $\Lambda_m^o$ is the limiting molar conductivity of the ions at a particular temperature.

Using the $\Lambda_m^o$ values at the different temperatures[1], we obtained the diffusion coefficient values of Na$^+$ (Table S3) as well as for Cl$^-$ (Figure S10). With the help of Stokes-Einstein relation which relates the diffusion coefficient and mobility, μ, by $D = \mu k_B T/e$, the mobility was calculated, where $k_B$ is the Boltzmann constant, and e is the elementary charge.

**Table S3.** The limiting molar conductivity, $\Lambda_m^o$, diffusion coefficient, D, and, mobility of bulk Na$^+$ at various temperatures.

| Temperature (K) | Limiting molar conductivity, $\Lambda_m^o$ ($10^{-4}$ m$^2$ S mol$^{-1}$) | Diffusion coefficient, D ($10^{-9}$ m$^2$ s$^{-1}$) | Mobility, μ ($10^{-8}$ m$^2$ V$^{-1}$ s$^{-1}$) |
|---|---|---|---|
| 298 | 50.14 | 1.33 | 5.19 |
| 303 | 55.67 | 1.50 | 5.76 |
| 308 | 61.49 | 1.69 | 6.37 |
| 313 | 67.49 | 1.88 | 6.99 |
| 318 | 73.65 | 2.09 | 7.63 |
| 323 | 79.89 | 2.30 | 8.28 |
| 328 | 86.56 | 2.53 | 8.97 |
| 333 | 93.26 | 2.77 | 9.66 |

The mobility of cations and anions, individually, can be calculated using the Henderson equation for mobility ratio:

$$\frac{\mu_+}{\mu_-} = -\left(\frac{z_+}{z_-}\right)\left(\frac{\ln(\Delta) - z_- F V_{diff}/RT}{\ln(\Delta) - z_+ F V_{diff}/RT}\right) \quad (2)$$

Where $\mu_+$ and $\mu_-$ are the cation and anion mobility respectively, $z_+$= +1 and $z_-$ = -1. To obtain $\mu_+$ and, $\mu_-$ we measured the conductivity of 1 M NaCl across the temperature range, 296 K to 333 K. Since conductivity can be described by: $\sigma \approx F(C_+\mu_+ + C_-\mu_-)$, where $C_+$ and $C_-$ are the concentrations of cations and anions, respectively. Using this, along with equation (2), will provide us with the mobility of cations and anions (Figure 4B, Figure S10).

**Note S6. COMSOL Multiphysics simulations.**

The enhanced osmotic power density with temperature through vermiculite membrane was modelled using a continuum based two dimensional coupled Poisson-Nernst-Planck equation using COMSOL Multiphysics[2] software. The Poisson equation describes the distribution of electric potential and the Nernst-Planck equation describes the concentration of cationic and anionic distribution through the channel.

The total ionic flux $J_\pm$ of each ionic species resulting from the contributions of the diffusion term associated with the concentration gradient and electrophoretic term associated with the electric potential gradient is shown below.

$$\nabla . J_\pm = \nabla . (-D_\pm \nabla C_\pm - z_\pm F \mu_\pm C_\pm \nabla \varphi) = 0 \tag{3}$$

Where $C_+$, $C_-$, $D_+$, $D_-$, $\mu_+$, $\mu_-$, $z_+$ and, $z_-$ are the concentrations, diffusion coefficients, ionic mobilities and valences of cations and anions, respectively and $F$ is the Faraday constant.

At high temperature, the ionic mobility follows the Arrhenius equation[3]

$$\mu_\pm = (\mu_{0\pm})exp\left(-\frac{E_a}{RT}\right)exp(\alpha N) \tag{4}$$

Where $\mu_{0\pm}$ is the Arrhenius constant, N = 2666 is the number of layers for a 4 μm thick membrane, $\alpha$ = 0.00487 is a fitting factor and $E_a$ = 36 kJ/mol is the activation energy.

Hence, according to Stokes-Einstein relation, the diffusion co-efficient of ion is

$$D_\pm = \frac{\mu_\pm k_B T}{e} \tag{5}$$

The total ionic current through the nanochannel is calculated by integrating the total ionic flux over one of the reservoirs $I = \int_S F(z_+ J_+ + z_- J_-) . n \, dS$

Where $n$ is the unit vector normal to the surface and S is the surface boundary.

An electrostatic potential arises from the space-dependent ion concentration as well as the fixed negative surface charge which is solved by using Poisson equation given as

$$-\nabla . (\varepsilon_r \nabla \varphi) = \frac{\rho_e}{\varepsilon_0} \tag{6}$$

Where $\varepsilon_0$ is the dielectric constant of free space, $\varepsilon_r$ is the relative dielectric constant of the aqueous solution and $\rho_e$ is the net space charge density of the ions, defined as

$$\rho_e = F(z_+ C_+ + z_- C_-) \tag{7}$$

In order to conserve the charge on the walls of the channel, the following electrostatic boundary condition is applied

$$\mathbf{n} . (\varepsilon_0 \varepsilon_r \nabla \varphi) = \sigma_s \tag{8}$$

Where $n$ is the unit normal vector pointing outwards to the nanochannel and $\sigma_s$ is the surface charge density which is taken to be -3.7 mC/m² calculated using Gouy-Chapman equation for a zeta potential of -45 mV (in 1 mM NaCl solution)[4].

The Poisson and NP equation are simultaneously solved in a 2D geometry with a nanochannel of height 0.5 nm connected to two reservoirs having dimensions (100 nm × 100 nm). The concentration on the permeate side is $C_L$ = 1 mM and on the feed side is $C_H$ = 100 mM. The boundary conditions are summarized below.

The boundary conditions for the PNP equation are

At the left reservoir, $\quad\quad\quad\quad\quad\quad C_\pm = C_H, \varphi = 0$ (9)
At the right reservoir, $\quad\quad\quad\quad\quad C_\pm = C_L, \varphi = V_{app}$ (10)

Here, we have taken a geometry which consists of a single sub-nanochannel of height, $h_{channel}$ = 0.5 nm and length, $L_{channel}$ = 150 nm, connected to two reservoirs (100 nm × 100 nm) on both the ends (Figure S11).

Cationic transfer number ($t_+$) and anionic transfer number ($t_-$) are calculated from the cationic current ($I_+$) and anionic current ($I_-$) at zero external electric potential using the following equations:

$$t_+ = \frac{|I_+|}{|I_+| + |I_-|}, 0 < t_+ < 1 \tag{11}$$

$$t_- = \frac{|I_-|}{|I_+| + |I_-|}, \; 0 < t_- < 1$$

(12)

The selectivity of the channel can then be calculated as:

$$S = t_+ - t_-$$

The selectivity is found out to be 0.91 for all the temperatures.

Power can be calculated from the diffusion potential and the current values extracted from Figure S12.

At T = 296 K, $V_{diff}$ = 96.14 mV, $I_{diff}$ = -1.25 × 10$^{-11}$ A

At T = 333 K, $V_{diff}$ = 108 mV, $I_{diff}$ = -6.3 × 10$^{-11}$ A

With Power = $I_{diff} \times V_{diff}$, the percentage change in the power density at 296 K and 333 K is estimated to be, 466 % which is marginally close to our experimental results.

**Table S4.** Parameters used in COMSOL simulations.

| Parameters | Symbols | Values |
| --- | --- | --- |
| Bulk diffusion co-efficient of Na$^+$ | $D_{Na^+}$ | 1.33 × 10$^{-9}$ m²/s |
| Bulk diffusion co-efficient of Cl$^-$ | $D_{Cl^-}$ | 2.03 × 10$^{-9}$ m²/s |
| Relative permittivity | $\varepsilon_r$ | 80 |
| Permittivity of the air | $\varepsilon_0$ | 8.854 x 10$^{-12}$ F/m |
| Faraday constant | F | 96,485 C/mol |
| Magnitude of Charge of electron | e | 1.6 x 10$^{-19}$ C |
| Valance of Na$^+$ | $z_+$ | +1 |
| Valance of Cl$^-$ | $z_-$ | -1 |
| Surface charge density | $\sigma_s$ | -3.7 mC/m² |
| Applied electric potential | $V_{app}$ | -0.2 V to 0.2 V |
| Gas constant | R | 8.314462 J/K |
| High concentration | $C_H$ | 100 mM |
| Low concentration | $C_L$ | 1 mM |
| Activation energy | $E_a$ | 36 kJ/mol |
| Layer number | N | 2666 |
| Fitting parameter | $\alpha$ | 0.00487 |

**SUPPLEMENTAL FIGURES**

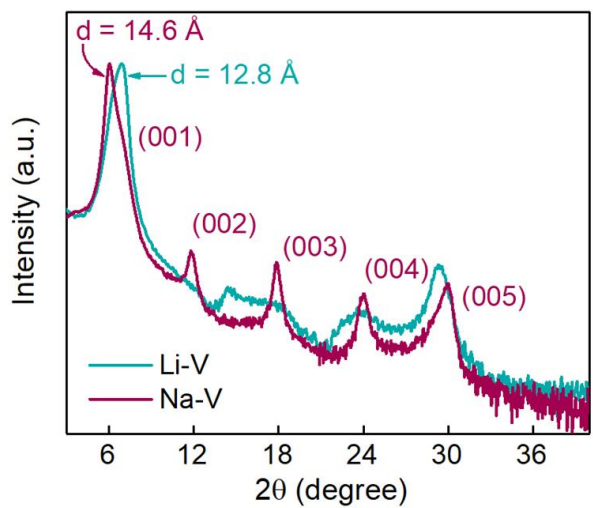

**Figure S1. XRD data of Li-V and Na-V membranes.** The XRD data shows the change in the d-spacing of Li-V after dipping in NaCl for 24 hours. The increased d-spacing and crystallinity of Na-V is evident from this plot.

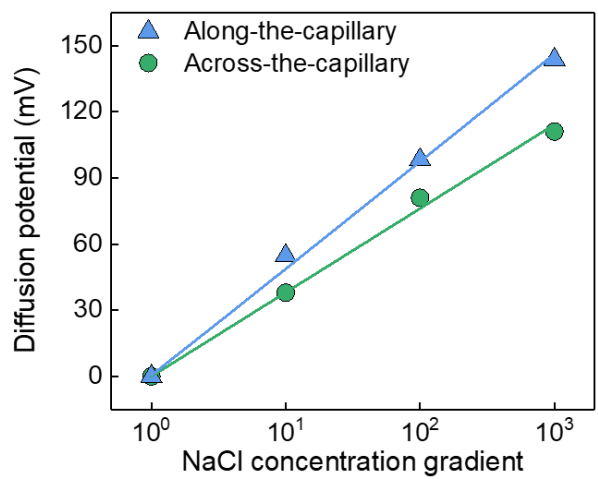

**Figure S2. Cation selectivity of Na-V membrane.** The plot of diffusion potential as a function of NaCl concentration gradient. The average cation selectivity was calculated using the Nernst equation for several concentration gradients. The selectivity calculated for 'along-the-capillary' and 'across-the-capillary' was 0.83 and 0.64, respectively. The solid lines represent the fit using Nernst equation.

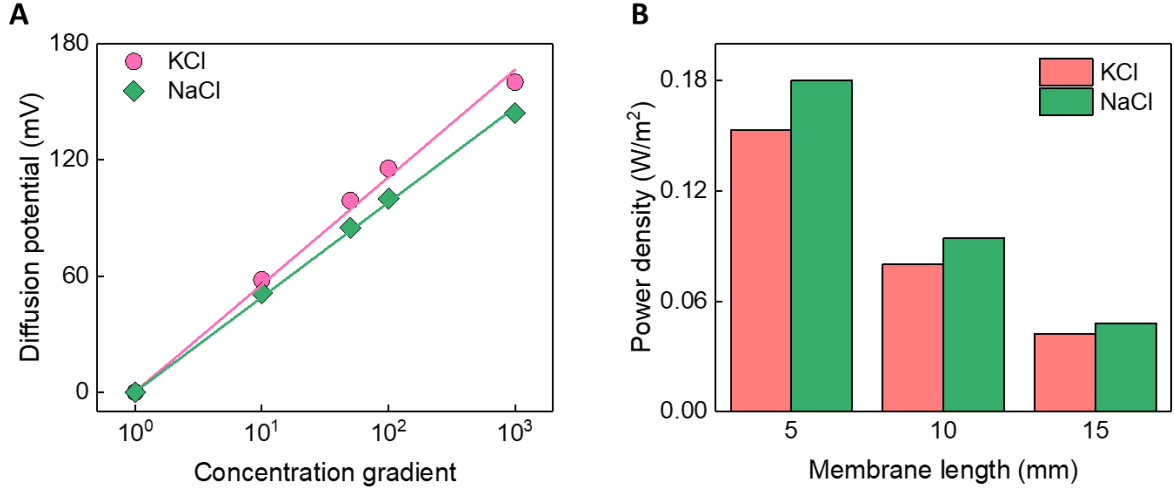

**Figure S3. Selectivity and power density comparison of K-V and Na-V membranes.** (A) Variation of diffusion potential as a function of KCl and NaCl concentration gradients via K-V and Na-V membranes (L ≈ 5 mm), respectively. The solid lines represent the fit using the Nernst equation. (B) Power density with KCl and NaCl using K-V and Na-V membranes, respectively, with lengths of 5 mm, 10 mm, and 15 mm.

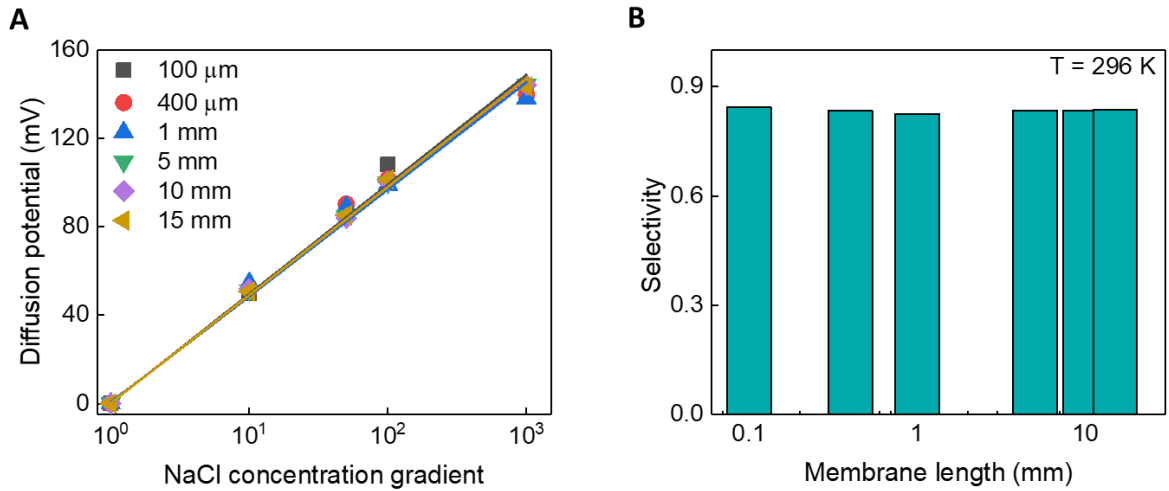

**Figure S4. Selectivity of Na-V membrane for different length at T = 296 K.** (A) Diffusion potential as a function of concentration gradient for membrane lengths ranging from 100 μm to 15 mm. The solid lines represent the fit using Nernst equation. (B) Selectivity as a function of membrane length.

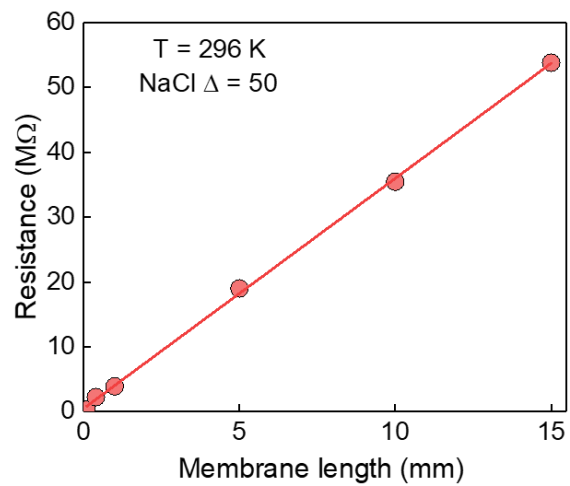

**Figure S5. Membrane resistance with length.** Resistance of the membrane increases linearly with increase in length. Here the length varies from 100 µm to 15 mm. The solid line represents the linear fit to the data.

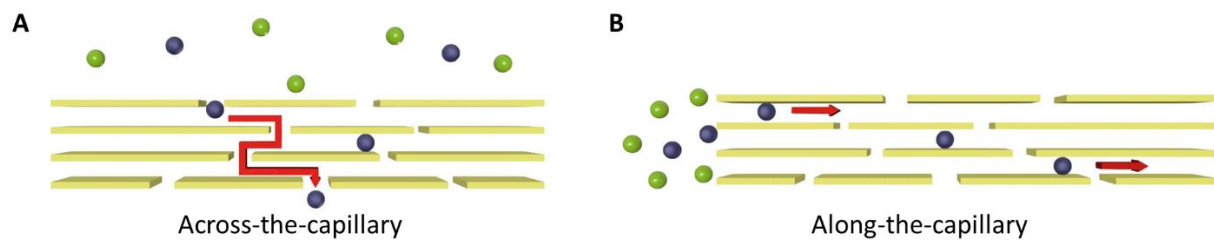

**Figure S6. Schematic of ion transport orientation.** (A) 'Across-the-capillary' transport of cations (blue) via a cation selective Na-V membrane and rejection of anions (green). (B) 'Along-the-capillary' transport of cations via a cation selective Na-V membrane.

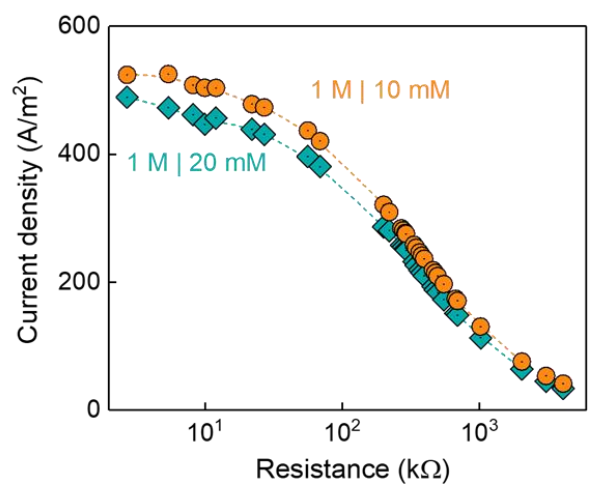

**Figure S7. Current density across Na-V membrane as a function of external load resistance.** The variation in current density for NaCl Δ = 50 and Δ = 100 at T = 296 K when the external load resistance is varied.

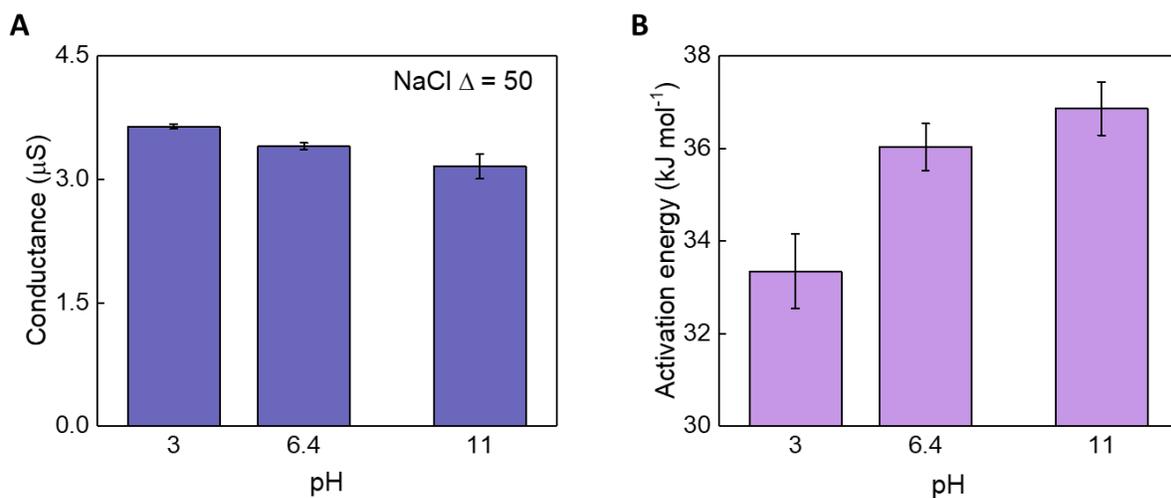

**Figure S8. Membrane's behavior in pH solutions.** (A) Decrease in conductance of NaCl solution with increase in pH of the solution. (B) Membrane activation energy with pH of solution ranging from 3 to 11 at Δ = 50. The error bars represent the standard deviation in three independent measurements.

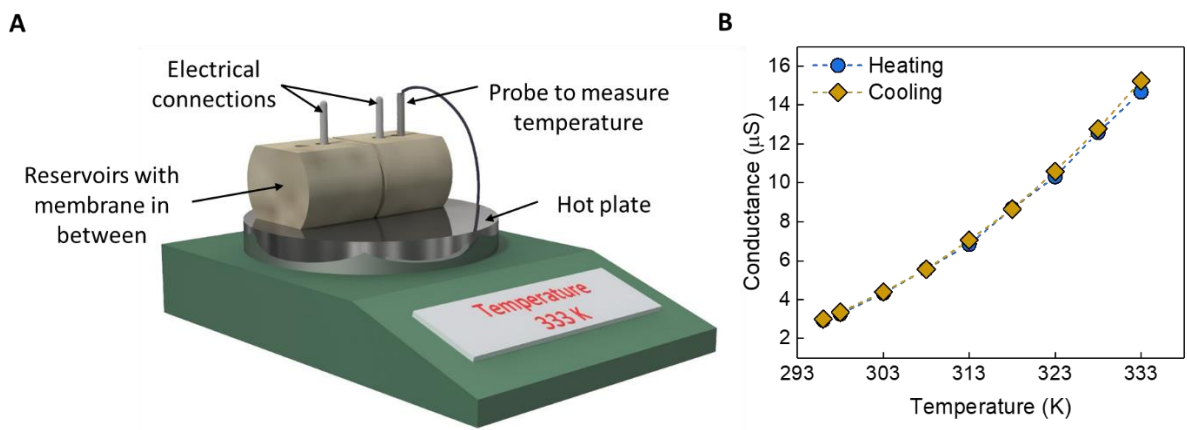

**Figure S9. Temperature dependence of conductance and power density.** (A) Experimental setup for temperature dependent studies. (B) Change in ionic conductance of membrane during heating and cooling, showing the absence of hysteresis. The dotted lines are a guide for the eye.

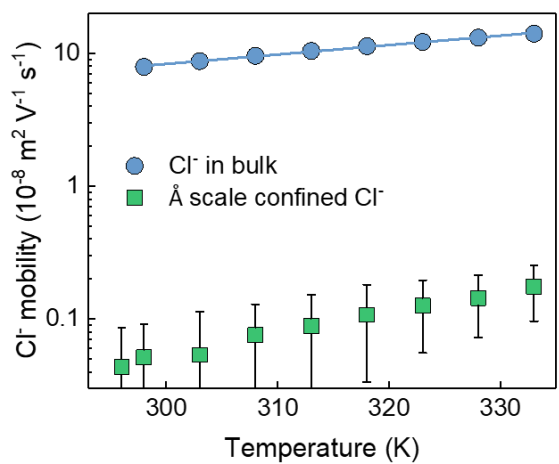

**Figure S10. Mobility of Cl⁻ through angstrom-scale channels of Na-V membrane.** The anions display a very low mobility as compared to bulk Cl⁻ as well as the cations (Figure 4B). The blue solid line represents the linear fit and the error bars indicate the standard deviation in repetitive measurements.

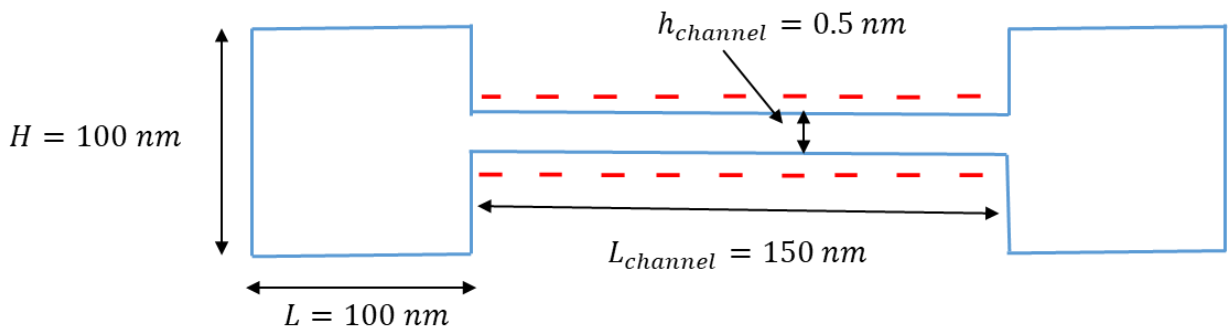

**Figure S11. A schematic illustrating nanochannel's geometric configuration,** where red dotted lines represent the negative surface charge (Figure not to scale). $H$, and $L$, respectively, are the height and length of both the reservoirs.

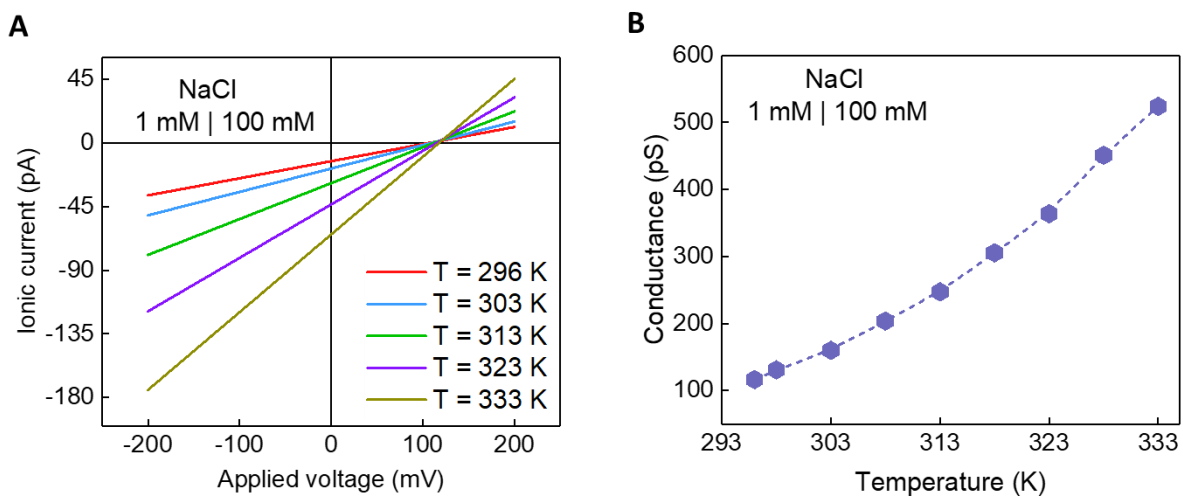

**Figure S12. COMSOL simulation of diffusion measurements for NaCl Δ = 100 at different temperatures.** (A) *I-V* characteristics of a single nanochannel with height 0.5 nm under a NaCl concentration gradient of 100. (B) Conductance as an exponentially increasing function of temperature over a range of 296 K to 333 K.

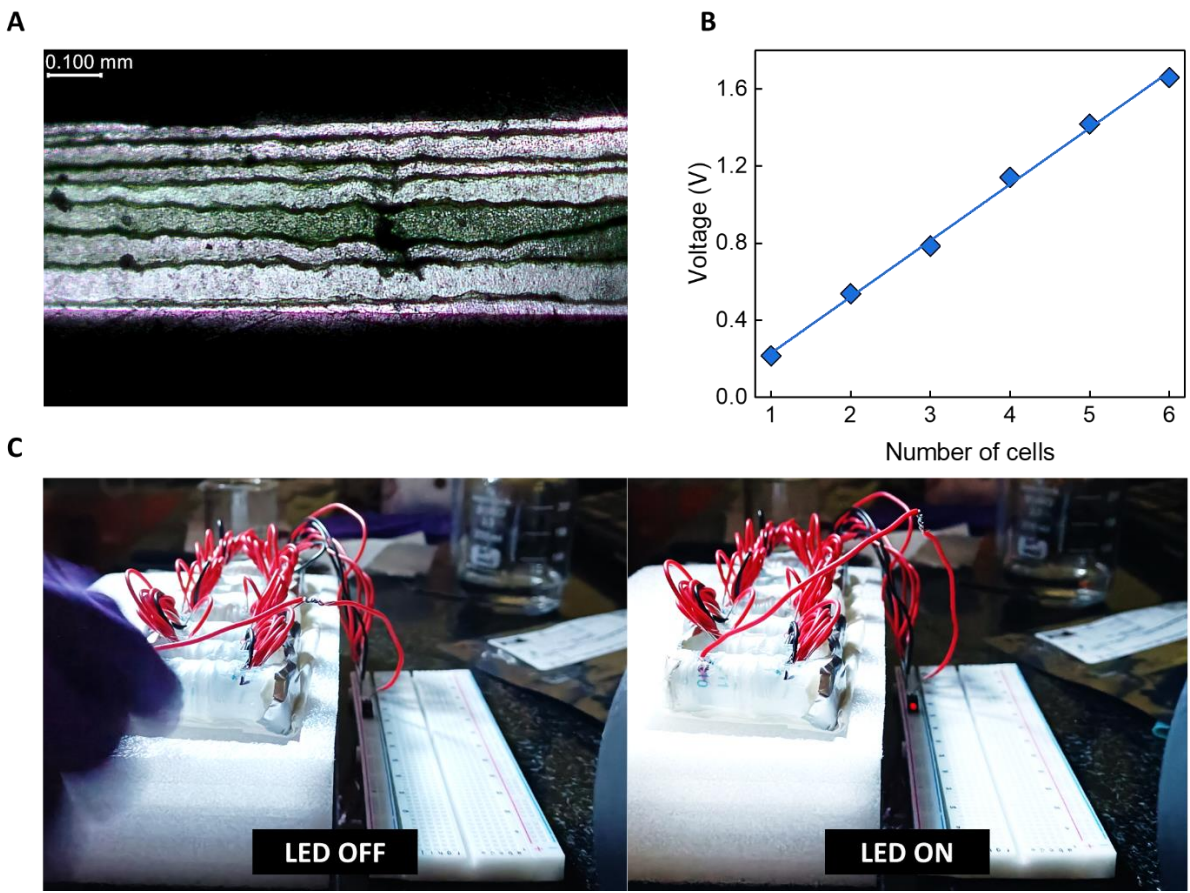

**Figure S13. Powering light emitting diodes and demonstration.** (A) Optical image of 8 membranes (L > 500 µm) stacked on top of each other to increase the effective area. (B) 6 cells with Δ = 150 connected in series showing a linear increase in zero-current potential with number of cells. The solid blue line represents the linear fit. (C) Several stacked membranes on sample holder placed between reservoirs and connected in series, showing a 3 mm Red LED (Dialight 5511109F) in OFF state (left) and in ON state (right).

**Table S5.** Power density comparison with several membranes used to harvest osmotic energy by using a NaCl concentration gradient of 50.

| Membrane | Power density (W/m$^2$) | Membrane length or thickness (μm) | Area (μm$^2$) | Surface charge or zeta potential |
|---|---|---|---|---|
| At room temperature | | | | |
| Heterogeneous polyelectrolyte/ Alumina nanochannel[5] | 5.13 | 1 | 3 × 10$^4$ | 80 mC/m$^2$ |
| Zwitterionic gradient double-network hydrogel membranes[6] | 5.44 | 360 | 3 × 10$^4$ | -60 mC/m$^2$ |
| bsGOM[7] | 4.9 | 4 | 3 × 10$^4$ | -34 mV |
| Mesoporous carbon-silica/anodized aluminum (MCS/AAO)[8] | 5.04 | 0.42 | 3 × 10$^4$ | +50 mC/m$^2$ |
| Asymmetric BAN [9] | 2.1 | 72.5 | 3 × 10$^4$ | -50 mC/m$^2$ |
| TFP-TPA COF@AFM[10] | 5.41 | 0.11 | 3 × 10$^4$ | -60 mC/m$^2$ |
| Ti$_3$C$_2$T$_X$ MXene[11] | 4.6 | 2 x 10$^3$ | 3 × 10$^4$ | -20 mC/m$^2$ |
| Ti$_3$C$_2$T$_X$ MXene-Based ionic diode membranes[12] | 8.6 | 4 | 3 × 10$^4$ | 0.5 mC/m$^2$ |
| Janus [13] | 2.04 | 0.5 | 3 × 10$^4$ | -80 mC/m$^2$ |
| Silk-based hybrid membranes[14] | 2.86 | 10 | 10$^8$ | -6 mC/m$^2$ |
| MXene/GO[15] | 3.7 | 0.42 | 7.1 x 10$^8$ | -36 mV |
| PCS-TOEC[16] | 2.98 | 10$^3$ | 9 x 10$^4$ | -60 mC/m$^2$ |
| Porous VMT[17] | 4.1 | 2.1 | 3 x 10$^4$ | -31 mV |
| GO/CNF[18] | 4.19 | 9 | 3 × 10$^4$ | – |
| MXene/Kevlar[19] | 3.7 | 2 | $3 \times 10^4$ | -20 mC/m² |
| Vertical-GO[20] | 10.6 | 350 | $1.5 \times 10^3$ | −73.8 mC/m² |
| "Along-the-capillary" vermiculite | 9.6 | 100 | $5.28 \times 10^2$ | -3.7 mC/m² |
| **At temperature ≥ 333 K** | | | | |
| PCS-TOEC [16] (335 K) | 7.7 | $10^3$ | $9 \times 10^4$ | -60 mC/m² |
| MXene/GO[15] (343 K) | 7.8 | 0.42 | $7.1 \times 10^8$ | -36 mV |
| "Along-the-capillary" vermiculite (333 K) | 65.1 | 100 | $5.28 \times 10^2$ | -3.7 mC/m² |